\documentstyle[12pt,epsfig]{article}

\newcommand{\re}{\mathop{\rm Re}\nolimits}
\newcommand{\im}{\mathop{\rm Im}\nolimits}

\catcode`@=11
\newcount\@tempcntc
\def\@citex[#1]#2{\if@filesw\immediate\write\@auxout{\string\citation{#2}}\fi
  \@tempcnta\z@\@tempcntb\m@ne\def\@citea{}\@cite{\@for\@citeb:=#2\do
    {\@ifundefined
       {b@\@citeb}{\@citeo\@tempcntb\m@ne\@citea\def\@citea{,}{\bf ?}\@warning
       {Citation `\@citeb' on page \thepage \space undefined}}%
    {\setbox\z@\hbox{\global\@tempcntc0\csname b@\@citeb\endcsname\relax}%
     \ifnum\@tempcntc=\z@ \@citeo\@tempcntb\m@ne
       \@citea\def\@citea{,}\hbox{\csname b@\@citeb\endcsname}%
     \else
      \advance\@tempcntb\@ne
      \ifnum\@tempcntb=\@tempcntc
      \else\advance\@tempcntb\m@ne\@citeo
      \@tempcnta\@tempcntc\@tempcntb\@tempcntc\fi\fi}}\@citeo}{#1}}
\def\@citeo{\ifnum\@tempcnta>\@tempcntb\else\@citea\def\@citea{,}%
  \ifnum\@tempcnta=\@tempcntb\the\@tempcnta\else
   {\advance\@tempcnta\@ne\ifnum\@tempcnta=\@tempcntb \else \def\@citea{--}\fi
    \advance\@tempcnta\m@ne\the\@tempcnta\@citea\the\@tempcntb}\fi\fi}
\catcode`@=12

\begin{document}

\title{\vskip-3cm{\baselineskip14pt
\centerline{\normalsize\hfill MPI/PhT/98--056}
\centerline{\normalsize\hfill hep--ph/9807545}
\centerline{\normalsize\hfill July 1998}}
\vskip1.5cm
Mass and Width of a Heavy Higgs Boson}
\author{{\sc Bernd A. Kniehl and Alberto Sirlin}\thanks{Permanent address:
Department of Physics, New York University, 4 Washington Place, New York,
NY~10003, USA.}\\
{\normalsize Max-Planck-Institut f\"ur Physik (Werner-Heisenberg-Institut),}\\
{\normalsize F\"ohringer Ring~6, 80805 Munich, Germany}}

\date{}

\maketitle

\thispagestyle{empty}

\begin{abstract}
The gauge dependence of the Higgs-boson mass and width in the on-shell scheme
of renormalization is studied in the heavy-Higgs-boson approximation.
The corresponding expansions in the pole scheme are analyzed adopting three 
frequently employed parametrizations.
The convergence properties and other theoretical features of the on-shell and
pole expansions, as well as their relative merits, are discussed.

\medskip

\end{abstract}

\newpage

There exists a significant and interesting literature concerning the analysis 
of the mass and width of a heavy Higgs boson, both in the pole and on-shell
schemes of renormalization \cite{mar,wil,ghi,fri,bin}.
The theoretical results may be conveniently expressed as expansions in
$\lambda/(2\pi)=GM_H^2$, where $M_H$ and $\lambda$ are the mass and quartic
coupling of the Higgs boson and $G=G_F/(2\pi\sqrt2)$.
Calling $g$ the SU(2) coupling, in the heavy-Higgs approximation (HHA), the 
limit $g,M_W,M_Z\to0$ with $G\propto g^2/M_W^2$ and $\lambda$ held fixed is 
employed, and the top-quark and other fermionic contributions are neglected.
In the HHA, the Higgs-boson width and the relation between the on-shell and 
pole masses are known through ${\cal O}(\lambda^3)$, i.e., in the 
next-to-next-to-leading order (NNLO) \cite{wil,ghi,fri,bin}.
Recently, however, it has been emphasized that, in the on-shell scheme, both 
the Higgs-boson mass and width are gauge-dependent quantities \cite{kni}.
In this letter, we re-examine the on-shell-scheme expansions in the HHA, with 
particular emphasis on the issue of gauge dependence.
We also re-analyze the pole-scheme expansions adopting three different,
frequently employed parametrizations, and discuss their convergence 
properties, as well as other theoretical features.

Calling $M_0$ the bare mass and $A(s)$ the self-energy, the on-shell mass $M$
and width $\Gamma$ of the Higgs boson are given by
\begin{equation}
M^2=M_0^2+\re A(M^2),\qquad
M\Gamma=-\frac{\im A(M^2)}{1-\re A^\prime(M^2)}.
\label{eq:os}
\end{equation}
Instead, in the pole scheme, one considers the complex-valued position of the
propagator's pole \cite{ede},
\begin{equation}
\bar s=M_0^2+A(\bar s).
\end{equation}
Given $\bar s$, there is no unique way to define the pole mass and width.
Two frequently employed parametrizations are
\begin{eqnarray}
\bar s&=&m_2^2-im_2\Gamma_2,\label{eq:p2}\\
\bar s&=&\left(m_3-\frac{i}{2}\Gamma_3\right)^2,
\label{eq:p3}
\end{eqnarray}
with $m_2$, $\Gamma_2$ or $m_3$, $\Gamma_3$ identified with the mass and width
of the unstable particle.
A third definition is
\begin{equation}
\bar s=\frac{m_1^2-im_1\Gamma_1}{1+\Gamma_1^2/m_1^2}
\label{eq:p1}
\end{equation}
or, equivalently,
\begin{equation}
m_1=\sqrt{m_2^2+\Gamma_2^2},\qquad
\Gamma_1=\frac{m_1}{m_2}\Gamma_2.
\label{eq:lep}
\end{equation}
In the $Z$-boson case, Eq.~(\ref{eq:lep}) leads, to very good approximation, 
to a Breit-Wigner resonance amplitude with an $s$-dependent width, and it has 
been shown that the $m_1$ definition can be identified with the $Z$-boson mass
measured at LEP \cite{sir}.
An important property of $\bar s$ and, therefore, also of
$m_i$ and $\Gamma_i$ ($i=1,2,3$), is that they are gauge-invariant quantities.
In order to elucidate the gauge dependence of $M$ and $\Gamma$, it is useful 
to compare Eq.~(\ref{eq:os}) with Eq.~(\ref{eq:p2}).
In the next-to-leading-order (NLO) approximation, one finds \cite{kni}
\begin{eqnarray}
\frac{M-m_2}{m_2}&=&-\frac{\Gamma_2}{2m_2}\im A^\prime(m_2^2)+{\cal O}(g^6),
\nonumber\\
\frac{\Gamma-\Gamma_2}{\Gamma_2}&=&
\im A^\prime(m_2^2)\left(\frac{\Gamma_2}{2m_2}+\im A^\prime(m_2^2)\right)
-\frac{m_2\Gamma_2}{2}\im A^{\prime\prime}(m_2^2)+{\cal O}(g^6),
\label{eq:lin}
\end{eqnarray}
where the prime indicates differentiation with respect to $s$.
In the Standard Model (SM), the one-loop bosonic contribution to $A(s)$ is
given by \cite{kni}
\begin{eqnarray}
\im A_{\rm bos}(s)&=&\frac{G}{4}s^2\left[-\left(1-\frac{4M_W^2}{s}
+\frac{12M_W^4}{s^2}\right)\left(1-\frac{4M_W^2}{s}\right)^{1/2}
\theta(s-4M_W^2)
\right.\nonumber\\
&&{}+\left.
\left(1-\frac{M^4}{s^2}\right)\left(1-\frac{4\xi_WM_W^2}{s}\right)^{1/2}
\theta(s-4\xi_WM_W^2)
+\frac{1}{2}(W\to Z)\right].
\label{eq:bos}
\end{eqnarray}
For finite values of the gauge parameters, $\xi_W$ and $\xi_Z$,
$\xi_WM_W^2,\xi_ZM_Z^2\to0$ as $M_W^2,M_Z^2\to0$.
Therefore, the second term contributes and cancels the leading $s$ dependence 
of the first one.
Thus, for finite values of $\xi_W$ and $\xi_Z$, one obtains in the HHA
\begin{equation}
\im A(s)=-\frac{3}{8}GM^4\qquad
\mbox{($R_\xi$ gauge)},
\label{eq:rxi}
\end{equation}
independent of $s$.
Denoting by $M_\xi$ and $\Gamma_\xi$ the on-shell mass and width in the 
$R_\xi$ gauge (defined for finite values of $\xi_W$ and $\xi_Z$) and applying 
henceforth the HHA, Eqs.~(\ref{eq:lin}) and (\ref{eq:rxi}) lead to
\begin{equation}
\frac{M_\xi}{m_2}=1+{\cal O}(\lambda^3),\qquad
\frac{\Gamma_\xi}{\Gamma_2}=1+{\cal O}(\lambda^3).
\label{eq:mrxi}
\end{equation}
Instead, in the unitary gauge, one first takes the limit
$\xi_W,\xi_Z\to\infty$, in which case the term proportional to
$\theta(s-4\xi_WM_W^2)$ in Eq.~(\ref{eq:bos}) does not contribute, and one 
finds
\begin{equation}
\im A(s)=-\frac{3}{8}Gs^2\qquad
\mbox{(unitary gauge)}.
\label{eq:uni}
\end{equation}
Noting that, in the unitary gauge, the one-loop expression for $\im A(s)$ 
involves couplings independent of $M$ for $s<4M^2$, Eq.~(\ref{eq:uni}) can 
independently be verified by replacing $M^2\to s$ in the well-known tree-level 
formula $M\Gamma=-\im A(M^2)=3GM^4/8$.
Denoting by $M_u$ and $\Gamma_u$ the on-shell quantities in the unitary gauge,
Eqs.~(\ref{eq:lin}) and (\ref{eq:uni}) tell us that
\begin{equation}
\frac{M_u}{m_2}=1+\frac{9}{64}G^2m_2^4+{\cal O}(\lambda^3),\qquad
\frac{\Gamma_u}{\Gamma_2}=1+\frac{9}{16}G^2m_2^4+{\cal O}(\lambda^3).
\label{eq:muni}
\end{equation}
Comparison of Eq.~(\ref{eq:mrxi}) with Eq.~(\ref{eq:muni}) shows that, in the 
HHA, the leading gauge dependence of the on-shell mass or width reduces to a 
discontinuous function, with one value corresponding to finite
$\xi_W$ and $\xi_Z$, and the other one to the unitary gauge.
It should be pointed out, however, that for finite and large values of $\xi_W$
and $\xi_Z$, the limit $\xi_WM_W^2,\xi_ZM_Z^2\to0$ is not realistic within the 
SM, and must be regarded as a special feature of the HHA.

The relation between $\Gamma_3$ and $m_3$ was first obtained in NNLO by 
Ghinculov and Binoth \cite{bin}, with a numerical evaluation of the expansion
coefficients.
We have independently derived this expansion.
The relation between $m_3$ and $M_\xi$ was first given analytically in NNLO by 
Willenbrock and Valencia \cite{wil}, an expansion that we have also verified.
As the connection between the three pole parametrizations $(m_1,\Gamma_1)$,
$(m_2,\Gamma_2)$, and $(m_3,\Gamma_3)$ is known exactly from
Eqs.~(\ref{eq:p2})--(\ref{eq:p1}), and the relations of $(M_\xi,\Gamma_\xi)$
and $(M_u,\Gamma_u)$ with $(m_2,\Gamma_2)$ are given to the required accuracy 
in Eqs.~(\ref{eq:mrxi}) and (\ref{eq:muni}), we readily find in NNLO the 
expansions of $\Gamma_i$ ($i=1,2,3,\xi,u$) in terms of $m_i$ in the three pole
and two on-shell schemes discussed above to be
\begin{equation}
\Gamma_i=\frac{3}{8}Gm_i^3\left[1+a\frac{Gm_i^2}{\pi}
+b_i\left(\frac{Gm_i^2}{\pi}\right)^2\right],
\label{eq:gam}
\end{equation}
where
\begin{eqnarray}
a&=&\frac{5}{4}\zeta(2)-\frac{3}{4}\pi\sqrt3+\frac{19}{8},\qquad
b_2=b_\xi=0.96923(13),\nonumber\\
b_1&=&b_2-\frac{9\pi^2}{64},\qquad
b_3=b_2-\frac{9\pi^2}{128},\qquad
b_u=b_2+\frac{9\pi^2}{64}.
\end{eqnarray}
For the ease of notation, we have put $m_\xi=M_\xi$ and $m_u=M_u$.
Here, we have adopted the value for $b_\xi$ from Ref.~\cite{fri}.
It slightly differs from the value $0.97103(48)$ determined in 
Refs.~\cite{ghi,bin}.
Although the difference is larger than the estimated errors, it amounts to 
less than 0.7\% in the coefficients $b_i$, which is unimportant for our
purposes.
On the other hand, $m_i$ ($i=1,3,\xi,u$) are related to $m_2$ by
\begin{equation}
m_i=m_2\left[1+c_i\left(\frac{Gm_2^2}{\pi}\right)^2
+d_i\left(\frac{Gm_2^2}{\pi}\right)^3\right],
\label{eq:mas}
\end{equation}
where
\begin{eqnarray}
c_1&=&\frac{9\pi^2}{128},\qquad
c_3=\frac{9\pi^2}{512},\qquad
c_\xi=0,\qquad
c_u=\frac{9\pi^2}{64},\nonumber\\
d_1&=&\frac{9\pi^2}{64}a,\qquad
d_3=\frac{9\pi^2}{256}a,\qquad
d_\xi=\frac{9\pi^2}{128}\left[-\frac{5}{4}\zeta(2)+\frac{\pi}{3}\sqrt3
+\frac{7}{8}\right],
\end{eqnarray}
while $d_u$ is currently unknown.
In the case of $i=3$, Eq.~(\ref{eq:gam}) agrees with Eq.~(10) of 
Ref.~\cite{bin} up to the numerical difference in $b_3$ discussed above.

\begin{figure}[ht]
\begin{center}
\epsfig{figure=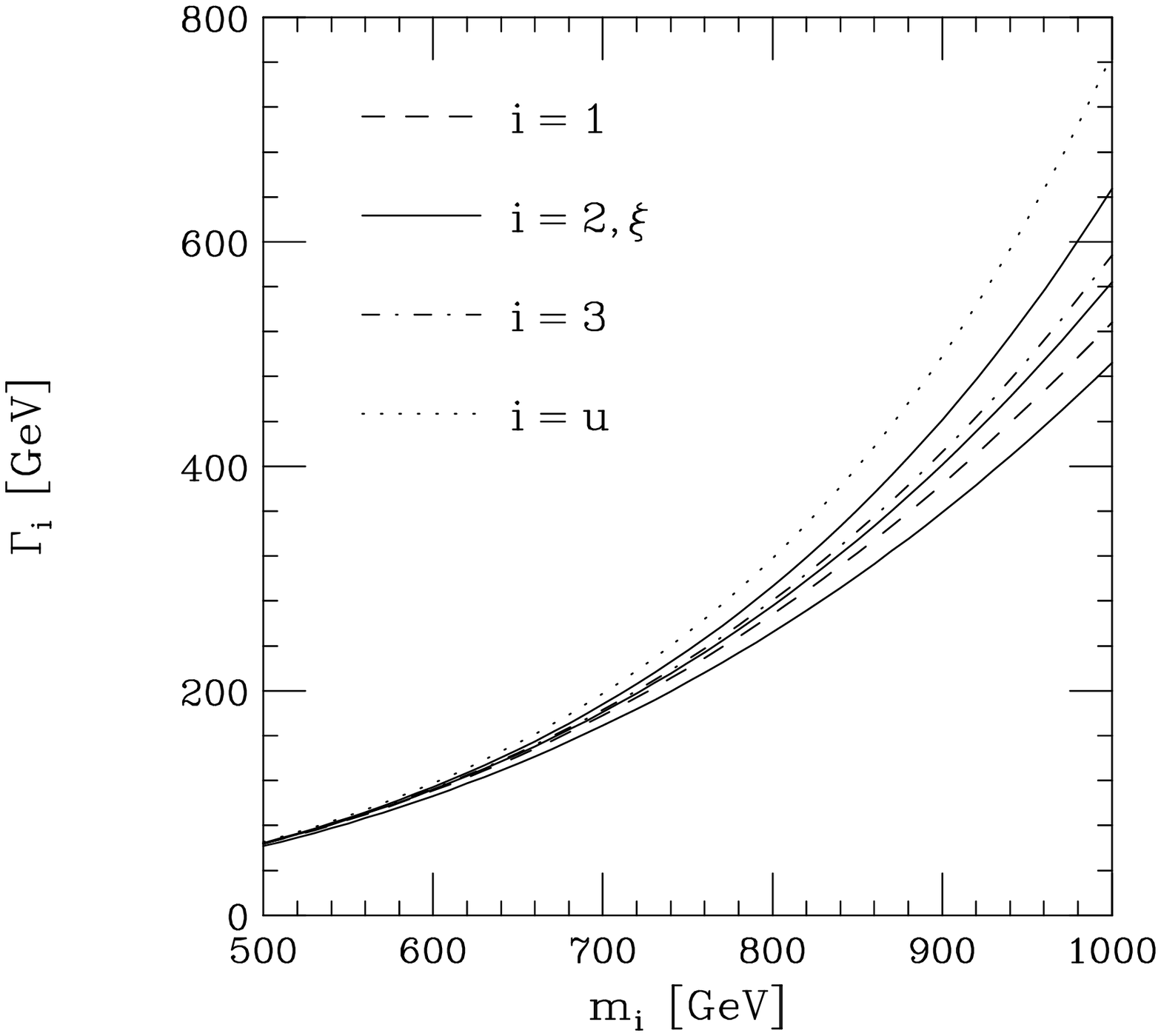,width=\textwidth}
\caption{Higgs-boson widths $\Gamma_i$ ($i=1,2,3,\xi,u$) as functions of the
corresponding masses $m_i$ in the various pole and on-shell schemes.
The down-most and middle solid lines correspond to the LO and NLO results, 
which are common to all renormalization schemes, while the up-most one refers
to the NNLO result for $i=2,\xi$.}
\label{fig:one}
\end{center}
\end{figure}

We see from Eq.~(\ref{eq:gam}) that all the width expansions have the same 
leading-order (LO) and NLO coefficients.
This is due to the fact that the on-shell and pole widths only differ in NNLO
\cite{kni} and that the relations (\ref{eq:mas}) among the masses do not
involve terms linear in $Gm_2^2/\pi$.
It is also interesting to note that the on-shell mass $M_{\rm PT}$ and width
$\Gamma_{\rm PT}$, defined in terms of the pinch-technique (PT) \cite{cor}
self-energy, obey Eq.~(\ref{eq:gam}) with $b_{\rm PT}=b_\xi$, and
Eq.~(\ref{eq:mas}) with $c_{\rm PT}=c_\xi=0$ \cite{kni}, while $d_{\rm PT}$ is
currently unknown.
In Fig.~\ref{fig:one}, the NNLO results for $\Gamma_i$ are plotted versus 
$m_i$ for the five cases considered in Eq.~(\ref{eq:gam}).
The down-most and middle solid curves depict the LO and NLO expansions,
respectively, which are common to the five cases.
The up-most solid curve corresponds to the NNLO expansion for $i=2,\xi$.
We note that $b_1$ is negative, while the other coefficients $b_i$ are
positive.
In particular, the NLO and NNLO corrections to $\Gamma_1(m_1)$ cancel at
$m_1=1.415$~TeV.

In order to analyze the scheme dependence of the above relations and the 
convergence properties of the corresponding perturbative series, one possible
approach \cite{bin} is to expand the relevant physical quantities in terms of
different masses $m_i$.
We illustrate this procedure with $m_2$ and $\Gamma_2$, which are the physical 
quantities that parametrize the conventional Breit-Wigner resonance amplitude,
proportional to $(s-m_2^2+im_2\Gamma_2)^{-1}$.
The relation $\Gamma_2(m_2)$ can be obtained directly from Eq.~(\ref{eq:gam}) 
or, via Eqs.~(\ref{eq:gam}) and (\ref{eq:mas}), from the expansions
\begin{eqnarray}
m_2&=&m_i\left[1-c_i\left(\frac{Gm_i^2}{\pi}\right)^2
-d_i\left(\frac{Gm_i^2}{\pi}\right)^3\right],
\label{eq:mtwo}\\
\Gamma_2&=&\frac{3}{8}Gm_i^3\left[1+a\frac{Gm_i^2}{\pi}
+(b_2-3c_i)\left(\frac{Gm_i^2}{\pi}\right)^2\right].
\label{eq:gtwo}
\end{eqnarray}
In the $m_i$-expansion scheme, for given $m_2$, one evaluates $m_i$ from 
Eq.~(\ref{eq:mtwo}) and $\Gamma_2$ from Eq.~(\ref{eq:gtwo}).
As the calculation of $\Gamma_2(m_i)$ through ${\cal O}(\lambda^n)$ only
requires the knowledge of $m_2(m_i)$ through ${\cal O}(\lambda^{n-1})$ and 
there is no term linear in $\lambda$ in Eq.~(\ref{eq:mtwo}), in LO (NLO), we
set $m_i=m_2$ and keep the first contribution (first and second contributions)
in Eq.~(\ref{eq:gtwo}), while in NNLO we retain the first two terms in
Eq.~(\ref{eq:mtwo}) and the three terms in Eq.~(\ref{eq:gtwo}).
In this manner, $m_2$ and $\Gamma_2$ are expanded to the same order in 
$\lambda$ relative to their respective Born approximations. 
Using as criterion of convergence the range throughout which the NNLO
corrections are smaller in magnitude than the NLO ones at fixed $m_2$, we find 
that the domains of convergence for the $m_1$, $m_2$, $m_3$, $M_\xi$, and
$M_u$ expansions are $m_2<733$~GeV, 930~GeV, 843~GeV, 930~GeV, and 672~GeV,
respectively.
In this connection, NLO (NNLO) correction means the difference between NLO and 
LO (NNLO and NLO) calculations.
We also find that these expansions, when restricted to the above ranges, are
in good agreement with each other.
Thus, the scheme dependence of the $\Gamma_2(m_2)$ relation is quite small
over the convergence domains of the expansions.

Another criterion that can be applied to judge the relative merits of the 
expansions is the closeness of the corresponding masses $m_i$ to $\bar m$, the 
peak position of the modulus of the $J=0$, iso-scalar Goldstone-boson
scattering amplitude.
The relation between $\bar m$ and $m_3$ is given to NNLO in Ref.~\cite{wil}.
Using Eq.~(\ref{eq:mas}), we can get the corresponding expressions for
$i=1,2,\xi$.
In the case of $i=2$, we have
\begin{equation}
m_2=\bar m\left[1+\frac{3\pi^2}{64}\left(\ln2-\frac{5}{2}\right)
\left(\frac{G\bar m^2}{\pi}\right)^2-0.778\left(\frac{G\bar m^2}{\pi}\right)^3
\right].
\end{equation}
An analytic expression for the NNLO coefficient is not available \cite{sco}.
For $\bar m=800$~GeV, we find
$m_1=0.984\,\bar m$, $m_2=0.925\,\bar m$, $m_3=0.940\,\bar m$, and
$M_\xi=0.934\,\bar m$, while, for $\bar m=1$~TeV, we have
$m_1=0.954\,\bar m$, $m_2=0.797\,\bar m$, $m_3=0.836\,\bar m$, and
$M_\xi=0.829\,\bar m$.

In summary, in this letter we have emphasized and explicitly exhibited the 
gauge dependence of the mass and width of a heavy Higgs boson in the on-shell
scheme.
We have also discussed the corresponding expansions in three frequently
employed parametrizations of the pole scheme.
Using our convergence criterion, the $m_i$ expansions, applied to the 
$\Gamma_2(m_2)$ relation, have domains of convergence with upper bounds
$(m_2)_{\rm max}$ in the range 672~GeV${}<(m_2)_{\rm max}<930$~GeV.
In this case, we find that the best convergence properties are exhibited by
the $m_2$ and $M_\xi$ expansions, followed in descending order by their $m_3$,
$m_1$, and $M_u$ counterparts.
We have also found that $m_1$ lies closest to the peak energy $\bar m$, 
followed by $m_3$ (the pole mass employed in Ref.~\cite{bin}), $M_\xi$, and
$m_2$.
Thus, from these considerations alone it is not possible to clearly establish 
the advantage of the pole schemes over their on-shell counterparts.
In our view, the fundamental importance of the pole-scheme expansions is that 
they involve gauge-invariant quantities, namely masses and widths that can be 
identified with physical quantities.

\vspace{1cm}
\noindent
{\bf Acknowledgements}
\smallskip

\noindent
We are grateful to Adrian Ghinculov for illuminating observations and to Kurt
Riesselmann for a private communication that helped us to evaluate the
coefficient $d_\xi$.
A.S. thanks the Theory Group of the Werner Heisenberg Institute for the
hospitality extended to him during a visit when this manuscript was prepared
and the Alexander von Humboldt Foundation for its kind support.
This research was supported in part by NSF Grant No.\ PHY--9722083.

\newpage

\end{document}